\documentclass[9pt,twocolumn,twoside]{opticajnl}
\journal{opticajournal} 

\setboolean{shortarticle}{false}
 
\usepackage{lineno}

\title{Orbital angular momentum control of third-harmonic generation and vortex dichroism in isotropic media}

\author[1]{Szymon Krukowski}
\author[2,*]{Kayn A. Forbes}

\affil[1]{School of Engineering, Mathematics, and Physics, University of East Anglia, Norwich Research Park, Norwich NR4 7TJ, United Kingdom}
\affil[2]{School of Chemistry, University of East Anglia, Norwich Research Park, Norwich NR4 7TJ, United Kingdom}

\affil[*]{k.forbes@uea.ac.uk}

\begin{abstract}
Structured light carrying orbital angular momentum enables new regimes of nonlinear light-matter interaction. Here we develop a molecular quantum electrodynamics description of third-harmonic generation (THG) driven by focused Laguerre-Gaussian beams in isotropic molecular media. We show that the nonparaxial longitudinal field components of a tightly focused beam permit THG with circularly polarized excitation in an isotropic fluid, a process forbidden for plane waves and paraxial beams. Within the electric-dipole approximation, the resulting emission is independent of the sign of the vortex charge. Including electric-magnetic dipole interference introduces a chiral contribution to the nonlinear response, giving rise to third-harmonic vortex dichroism (THVD). The emitted intensity then acquires a component linear in the topological charge \(\ell\), reversing sign with either the wavefront handedness or molecular chirality. Numerical modelling reveals corresponding spatial asymmetries in the harmonic field. These results establish both an allowed pathway for circularly polarized THG in isotropic fluids and the first chiroptical analogue of THG in such media, identifying orbital angular momentum as a new control parameter for nonlinear chiral spectroscopy.
\end{abstract}

\setboolean{displaycopyright}{false} 

\begin{document}

\maketitle

\section{Introduction}
Nonlinear optics studies the interaction of intense electromagnetic fields with matter, where the induced material response becomes nonlinear in the applied optical field \cite{gu2016molecular, zyss2013molecular, andrews2002optical}. Such effects typically arise at the high field strengths produced by lasers and give rise to a wide range of phenomena, including harmonic generation, wave mixing, and nonlinear scattering processes \cite{craig1998molecular, andrews2002optical}. These interactions form the basis of many modern spectroscopic and imaging techniques \cite{zipfel2003nonlinear, yu2008molecular, smirnova2019third, bonacina2020harmonic, xu2024multiphoton, potma2024foundations}, and continue to provide new opportunities for controlling light-matter interactions \cite{faucher2016rotational, bradshaw2019off}.

In recent years there has been increasing interest in the role of the spatial structure of light in nonlinear optical processes \cite{buono2022nonlinear, lou2022third}. Structured light fields, whose amplitude, phase, and polarization distributions are deliberately engineered \cite{forbes2021structured, he2022towards}, provide additional degrees of freedom for tailoring optical interactions. Of particular relevance in this work are optical vortex beams, which possess a helical phase structure of the form $\exp(i\ell\phi)$ and carry $\ell\hbar$ units of orbital angular momentum (OAM) per photon \cite{andrews2012angular}. The unique properties of these beams have led to numerous applications in optical manipulation, communication, and quantum optics \cite{shen2019optical}.

A molecule is said to be chiral if it cannot be superposed onto its mirror image by any combination of rotations and translations, and chirality plays a central role throughout chemistry. Chiral molecules therefore exist as pairs of enantiomers with identical chemical composition but opposite handedness. These enantiomers can interact differently with other chiral systems, leading to discriminatory behaviour in chemical reactions and molecular interactions. This selectivity underpins much of biochemistry and drug efficacy, making the assignment of absolute handedness a critical capability in health and pharmaceutical applications. 

Light may also be chiral, the handedness arising from both its polarization and spatial phase structure \cite{forbes2026twisted, forbes2026vortex}. Optical activity arises when such chiral molecules interact differently with left- and right-handed light, a phenomenon that underpins chiroptical spectroscopy and enables the determination of absolute molecular configuration \cite{polavarapu2016chiroptical, barron2009molecular}. The vast majority of such techniques rely on the chirality associated with circularly polarized states of light.

The topological charge $\ell$ of a vortex beam not only determines its OAM but, as a pseudoscalar, also defines the handedness of its wavefront. This introduces a distinct form of optical chirality associated with the sign of $\ell$. Unlike the bounded polarization helicity $\sigma\in[-1,1]$, the wavefront helicity $\ell\in\mathbb{Z}$ is unbounded, offering new routes to enhanced chiral interactions. Consequently, interactions between twisted light and chiral matter have attracted considerable interest. In particular, vortex dichroism describes a differential optical response to light carrying opposite signs of orbital angular momentum \cite{forbes2026twisted, forbes2026vortex}. Such effects arise from the coupling between the handedness of the optical wavefront and the intrinsic chirality of matter.

Despite growing interest in structured light, the role of orbital angular momentum in nonlinear chiroptical processes remains largely unexplored \cite{forbes2020nonlinear, mayer2024chiral, cheeseman2025nonlinear}. In particular, third-harmonic generation (THG) in isotropic molecular media is typically forbidden for circularly polarized excitation within the electric-dipole approximation \cite{andrews2002optical}, and no direct nonlinear analogue of vortex dichroism has previously been established.

In this work we investigate nonlinear interactions of focused optical vortex beams in isotropic molecular media. Within a molecular quantum electrodynamics framework, we show that third-harmonic generation with circularly polarized input is not forbidden. We then introduce a new nonlinear chiroptical effect, \emph{third-harmonic vortex dichroism} (THVD), and demonstrate how the handedness of optical vortices gives rise to chiral-sensitive nonlinear optical signals through the spatial structure of the generated fields, establishing a new route to nonlinear chiral spectroscopy.

\section{Third-Harmonic Generation with Laguerre-Gaussian modes}
In the theory of molecular quantum electrodynamics (MQED) \cite{craig1998molecular}, the interaction Hamiltonian for the dipole coupling of some molecule $\xi$ with a radiation field can be written as
\begin{align}
    H_{\textup{int}}(\xi)=-\boldsymbol{\mu}(\xi)\cdot \textbf{E}(\textbf{R}_\xi) -\textbf{m}(\xi)\cdot \textbf{B}(\textbf{R}_\xi),
    \label{eq:Hint1}
\end{align}
where $\boldsymbol{\mu}(\xi)$ and $\textbf{m}(\xi)$ are the electric and magnetic transition dipole moment operators, and $\textbf{R}_\xi$ is the position vector of $\xi$. The electric and magnetic field operator mode expansions for a Laguerre-Gaussian (LG) beam propagating in the $z$-direction, in cylindrical coordinates $(r,\phi,z)$, can be expressed as \cite{forbes2021relevance}:
\begin{align}
    \mathbf{E}_\text{LG}(\mathbf{r})=&\sum_{k,\ell,p} {\Omega} \biggl[\Bigl(\alpha \mathbf{\hat{x}}+\beta \mathbf{\hat{y}}+\mathbf{\hat{z}}\frac{i}{k}\Bigl\{\alpha\Bigl(\gamma \cos\phi - \frac{i \ell }{r} \sin\phi\Bigr) \nonumber \\  + &\beta \Bigl(\gamma \sin\phi + \frac{i \ell }{r} \cos\phi\Bigr) \Bigr\} \Bigr) \nonumber \\ & \times f_{\mathrm{LG}}\; {a}_{\ell,p}\; \mathrm{e}^{i(kz+\ell\phi)} - \mathrm{H.c}\biggr]
    \label{eq:ELG}
\end{align}
and
\begin{align}
    \mathbf{B}_{\mathrm{\text{LG}}}(\mathbf{r})=&\sum_{k,\ell,p} \frac{{\Omega}}{c} \biggl[\Bigl(\alpha \mathbf{\hat{y}}-\beta \mathbf{\hat{x}}+\mathbf{\hat{z}}\frac{i}{k}\Bigl\{\alpha\Bigl(\gamma \sin\phi + \frac{i \ell }{r} \cos\phi\Bigr)\nonumber \\ & -\beta \Bigl(\gamma \cos\phi - \frac{i \ell }{r} \sin\phi\Bigr) \Bigr\}\Bigr) \nonumber \\
    & \times f_{\mathrm{LG}}\; {a}_{\ell,p}\; \mathrm{e}^{i(kz+\ell\phi)} - \mathrm{H.c}\biggr]
    \label{eq:BLG}
\end{align}
where $\textbf{r}$ is a position vector; $k$ is the wavenumber; $\ell$ is the previously mentioned topological charge, where $\ell\in\mathbb{Z}$; $p$ is the radial index, where $p\in\mathbb{N}$; $\Omega=i\Big(\frac{\hbar ck}{2\varepsilon_0VA^2_{|\ell|,p}}\Big)^\frac{1}{2}$ is a normalization constant; $i$ is the imaginary unit; $\hbar$ is the reduced Planck's constant; $c$ is the speed of light; $\varepsilon_0$ is the vacuum permittivity; $V$ is the quantization volume; $A_{|\ell|,p}$ is a constant that ensures correct energy normalization of the LG mode, including both transverse and longitudinal field components \cite{forbes2021relevance}; $\alpha$ and $\beta$ are the complex Jones vector coefficients that must meet the requirement $|\alpha|^2+|\beta|^2=1$; $\gamma = \frac{|\ell|}{r}-\frac{2r}{w[z]^2}+ \frac{ikr}{R[z]} -\frac{4r}{w[z]^2}\frac{L^{|\ell|+1}_{p-1}}{L^{|\ell|}_{p}}$ \cite{forbes2026vortex, green2023optical}, where $w[z]$ denotes the beam width at some distance $z$; $R[z]$ is the radius of curvature of the beam's wavefronts at $z$; $L_p^{|\ell|}$ is the generalized Laguerre polynomial; $\phi$ is the azimuthal angle around the beam profile; $f_{\text{LG}}$ is the radial distribution function of the LG modes for some value of $z$, given by \cite{andrews2012angular}:
\begin{align}
    f_{\text{LG}} &= \sqrt{\frac{2p!}{{\pi w_{0}^2}(p+|\ell|)!}}\frac{w_0}{w[z]} \Biggr(\frac{\sqrt{2}r}{w[z]}\Biggr)^{|\ell|}L_{p}^{|\ell|}\Biggr[\frac{2r^2}{w^2[z]}\Biggr] \nonumber \\ & \times\textup{e}^{-\frac{r^2}{w^2[z]}}\textup{e}^{i\left(\frac{kr^2}{2R[z]} - (2p + |\ell| + 1)\zeta [z]\right)} \label{eq:fLG}
\end{align}
where $w_0$ is the beam waist; $\zeta[z]$ is the Gouy phase; $a_{\ell,p}$ is the annihilation operator; and H.c is the associated Hermitian conjugate, representing the terms associated with the raising operator $a^\dagger_{\ell,p}$. 

The $\mathbf{\hat{x}}$ and $\mathbf{\hat{y}}$-dependent terms in Eqs. (\ref{eq:ELG}-\ref{eq:BLG}) represent the 2D polarization state of the electric and magnetic fields. This transverse property of light (with respect to the direction of propagation, $z$) is well understood using paraxial models of light beams \cite{adams2018optics}. However, in order to satisfy Maxwell's equations, all real electromagnetic beams must contain longitudinal (non-paraxial) components.  The magnitude of the $\mathbf{\hat{z}}$-dependent terms in Eqs. (\ref{eq:ELG}-\ref{eq:BLG}) increase with a tighter focus relative to the transverse components. 

\begin{figure}[ht]
\centering
{\includegraphics[width=0.95\linewidth]{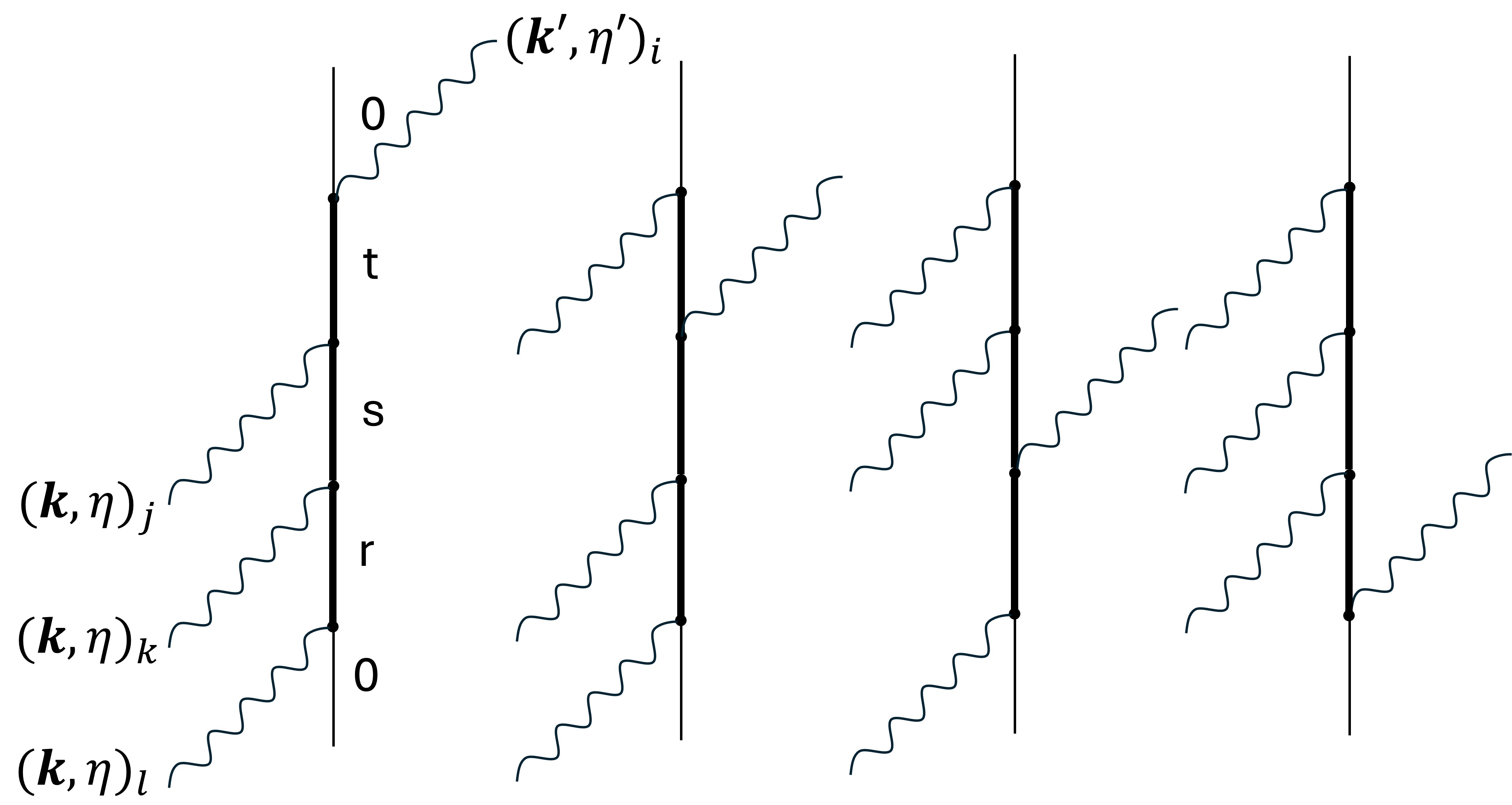}}
    \caption{\small{Four topologically distinct time-ordered Feynman diagrams describing THG required in the sum over all pathways. Time flows vertically. The intermediate states are denoted as $|r\rangle$, $|s\rangle$, and $|t\rangle$. Each photon is labelled using Latin indices, $i$, $j$, $k$, and $l$, additionally, they have an associated wave vector, $\textbf{k}$, and polarization constant, $\eta$.}}
    \label{fig:THG}
\end{figure}

THG is a four-photon process in which three input photons of frequency $\omega$ are annihilated, and a single photon of frequency $3\omega$ is created \cite{craig1998molecular, andrews1980harmonic}. The topologically distinct Feynman (time-ordered) diagrams required to describe THG are illustrated in Figure \ref{fig:THG}. The initial state of the light-molecule system may be written as a product wave function given by $|I\rangle=|n(\textbf{k},\eta)\rangle\prod\limits_\xi^N|E_0(\xi)\rangle$, where $n$ is the occupation number of the incident beam, $\eta$ designates the polarization state, $E_0(\xi)$ is the initial energy state of $\xi$, and $N$ is the number of molecules.

Under the assumption of purely electric-dipole interactions (first term in~\eqref{eq:Hint1}), third-order time-dependent perturbation theory yields the amplitude:
\begin{align}
    M_{FI}
    =& \sum\limits_{R,S,T} \frac{\langle F|H_{\textup{int}}|T\rangle\langle T|H_{\textup{int}}|S\rangle\langle S|H_{\textup{int}}|R\rangle\langle R|H_{\textup{int}}|I\rangle}{(E_I-E_R)(E_I-E_S)(E_I-E_T)} \nonumber\\
    =& -\sqrt{3}\bigg(\frac{\hbar ck}{2\varepsilon_0V}\bigg)^2\sqrt{n(n-1)(n-2)}\,
    \overline{e}_i'e_je_ke_l \nonumber \\ & \times
    \sum\limits_\xi \beta_{ijkl}(\xi)\,
    \overline{\tilde{f}'_{\mathrm{LG}}(r_\xi,z_\xi)}\,\tilde{f}_{\mathrm{LG}}^3(r_\xi,z_\xi)\,
    \mathrm{e}^{i\Delta\Phi(r_\xi,\phi_\xi,z_\xi)},
\end{align}
where the phase mismatch is 
\begin{align}
\Delta\Phi(r_\xi,\phi_\xi,z_\xi)
=&\;
(3k-k')z_\xi
+
(3\ell-\ell')\phi_\xi \nonumber\\
&+
\frac{r_\xi^2}{2}\left(\frac{3k}{R(z_\xi)}-\frac{k'}{R'(z_\xi)}\right)
\nonumber \\ &-
\left[
3(2p+|\ell|+1)\zeta(z_\xi)
-
(2p'+|\ell'|+1)\zeta'(z_\xi)
\right].
\label{phase}
\end{align}
\noindent Note that we separate the phase from the Laguerre-Gaussian mode function by writing
\begin{align}
f_{\mathrm{LG}}(r,z)
= \tilde{f}_{\mathrm{LG}}(r,z)\,
\exp\!\left[i\left(\frac{kr^2}{2R[z]}-(2p+|\ell|+1)\zeta[z]\right)\right],
\end{align}
where $\tilde{f}_{\mathrm{LG}}$ denotes the LG mode with the Gouy and wavefront curvature phases removed.

In the case of linearly polarized input fields, the nonlinear source phase is determined solely by the orbital contribution $3\ell\phi_\xi$. For circularly polarized input, an additional spin-dependent contribution modifies the azimuthal phase, leading to a source phase of the form $\mathrm{e}^{i(3kz+(3\ell+2\sigma)\phi)}$, and correspondingly altered phase-matching conditions. Both of these cases are considered explicitly in the following subsections. We employ Einstein summation notation throughout, with repeated indices implying summation. Here $e_i$ denotes the electric field polarization vector of photon $i$, and an overbar indicates complex conjugation. $E_I$ represents the total energy of the system in its initial state. The spatial mode function for the emitted photon, $f_{\mathrm{LG}}'$, is defined using~(\ref{eq:fLG}) with the substitution $k' \rightarrow 3k$, reflecting energy conservation in the third-harmonic process, while the azimuthal phase of the emitted mode is fixed by conservation of total angular momentum of the radiation field. For the linearly polarized case, energy and orbital-angular-momentum conservation imply $k'=3k$ and $\ell'=3\ell$, so that the longitudinal and azimuthal phase-matching conditions are satisfied. However, the Gouy phase introduces a mode-order-dependent axial mismatch away from the focal plane, limiting coherent buildup and reducing conversion efficiency. The fourth-rank hyperpolarizability tensor $\beta_{ijkl}$ contains the molecular response terms; its general form is provided in Supplementary Information. Here and throughout, Latin indices refer to laboratory-frame components, while Greek indices refer to molecular-frame components. For a homogeneous isotropic solution of identical molecules, the discrete sum over molecular centres may be replaced by a continuum integral with constant number density $\rho_0$, such that
\begin{align}
\sum_\xi \beta_{ijkl}(\xi)\,
\overline{\tilde{f}'_{\mathrm{LG}}(r_\xi,z_\xi)}\,\tilde{f}_{\mathrm{LG}}^3(r_\xi,z_\xi)\,
\mathrm{e}^{i\Delta\Phi(r_\xi,\phi_\xi,z_\xi)}
\;\nonumber \\ \rightarrow\;
\rho_0
\int_V d^3R\,
\beta_{ijkl}(\mathbf R)\,
\overline{\tilde{f}'_{\mathrm{LG}}(\mathbf R)}\,\tilde{f}_{\mathrm{LG}}^3(\mathbf R)\,
\mathrm{e}^{i\Delta\Phi(\mathbf R)}.
\end{align}
The THG rate follows from Fermi's golden rule,
\begin{align}
\Gamma=\frac{2\pi}{\hbar}\rho_f\,|M_{FI}|^2,
\end{align}
where $\rho_f$ is the density of final radiation states and $M_{FI}$ is the total scattering amplitude for the molecular ensemble. The corresponding radiant intensity follows as $I=2\hbar ck\,\frac{d\Gamma}{d\Omega}$ for an infinitesimal element of solid angle $d\Omega$ about the forward direction. The THG intensity scales cubically with the mean irradiance $\overline{I}_0$. Taking $\rho_f=\frac{(2k')^2 d\Omega}{(2\pi)^3\hbar c}V$, $\overline{I}_0=\frac{\langle n\rangle\hbar kc^2}{V}$, and defining the third-order coherence function $g^{(3)}=\frac{\langle n(n-1)(n-2)\rangle}{\langle n\rangle^3}$, the third-harmonic source intensity in the focal region may be written approximately as
\begin{align}
    I_\text{THG}=\frac{81\overline{I}_0^3g^{(3)}k^4N_{\mathrm{eff}}^{\,2}}{64\pi^2\varepsilon_0^4c^2}
    \tilde{f}^6_{\mathrm{LG}}\tilde{f}'^2_{\mathrm{LG}}
    \left|\overline{e}_i'e_je_ke_l\langle\beta_{ijkl}\rangle\right|^2,
    \label{eq:intensityNoAve}
\end{align}
where $N_{\mathrm{eff}}$ denotes an effective number of coherently contributing molecules within the interaction region. This expression therefore represents the local nonlinear source intensity rather than the fully accumulated third-harmonic output from an extended sample. In~\eqref{eq:intensityNoAve} the orientational average is taken within the modulus square because third-harmonic generation is a coherent process: the emitted fields from different molecules retain fixed phase relations, so the amplitudes add prior to squaring and interference terms must be preserved. By contrast, for incoherent processes phase correlations are randomized across emitters, the cross terms vanish on averaging, and the averaging is performed after squaring at the level of intensities.

~(\eqref{eq:intensityNoAve}) gives the local coherent source intensity prior to orientational averaging, for molecules with fixed orientation relative to the optical vortex beam. To account for randomly oriented molecules in an isotropic fluid, we perform a fourth-rank rotational average \cite{craig1998molecular} using standard methods (see Supplementary Information), yielding
\begin{align}
   I_\text{THG}  =& \frac{81\overline{I}_0^3g^{(3)}k^4N_{\mathrm{eff}}^{\,2}}{1600\pi^2\varepsilon_0^4c^2}
    \tilde{f}^6_{\mathrm{LG}}\tilde{f}'^2_{\mathrm{LG}}
    |\textbf{e}\cdot\textbf{e}|^2
    |\textbf{e}\cdot\overline{\textbf{e}}'|^2
    |\beta_{\lambda\lambda\mu\mu}|^2.
    \label{eq:intensityAnswer}
\end{align}

The spatial intensity distribution is obtained from the mode functions $\tilde{f}_{\mathrm{LG}}$ and $\tilde{f}'_{\mathrm{LG}}$, together with the corresponding vectorial field structure of the tightly focused vortex beams. It is noted that, while energy conservation ($k'=3k$) and angular momentum conservation determine the longitudinal and azimuthal phase relations of the emitted mode, the Gouy phase introduces a mode-order-dependent axial phase mismatch away from the focal plane. As such, Eqs.~(\ref{eq:intensityNoAve})-(\ref{eq:intensityAnswer}) are most rigorously interpreted as describing the local source intensity, with reduced conversion efficiency expected away from the focus due to incomplete phase matching.

The electric field vectors in~(\ref{eq:intensityAnswer}) are given by
\begin{align}
    \textbf{e}(\textbf{r})&=\alpha\hat{\textbf{x}}+\beta\hat{\textbf{y}}\nonumber \\ &+\hat{\textbf{z}}\frac{i}{k}\bigg\{\alpha\bigg(\gamma\cos\phi-\frac{i\ell}{r}\sin\phi\bigg)+\beta\bigg(\gamma\sin\phi+\frac{i\ell}{r}\cos\phi\bigg)\bigg\},
    \label{eq:8}
\end{align}
and
\begin{align}
    \overline{\textbf{e}}'(\textbf{r})&=\overline{\alpha}'\hat{\textbf{x}}+\overline{\beta}'\hat{\textbf{y}}\nonumber \\ &-\hat{\textbf{z}}\frac{i}{k'}\bigg\{\overline{\alpha}'\bigg(\overline{\gamma}'\cos\phi+\frac{i\ell'}{r}\sin\phi\bigg)+\overline{\beta}'\bigg(\overline{\gamma}'\sin\phi-\frac{i\ell'}{r}\cos\phi\bigg)\bigg\},
    \label{eq:9}
\end{align}
where $\alpha$ and $\beta$ define the three-dimensional polarization state of the incident field, and $\alpha'$ and $\beta'$ that of the emitted harmonic field. 

\subsection{Linear polarization}

We first consider THG with linearly polarized incident light. In this case the polarization coefficients $\alpha$ and $\beta$ are real, so the field carries no spin angular momentum. The spatial phase structure of the beam is determined solely by the orbital component $\text{e}^{i\ell\phi}$ associated with the vortex wavefront.

In the third-harmonic process the nonlinear source term contains the product of three incident fields, giving an overall phase dependence proportional to $\text{e}^{i3(kz+\ell\phi)}$ (along with the wavefront curvature and Gouy phase). Consequently, the emitted harmonic field must possess the longitudinal and azimuthal phase factors $\text{e}^{i3kz}$ and $\text{e}^{i3\ell\phi}$ in order to satisfy phase matching and conservation of angular momentum. The emitted vortex mode therefore has azimuthal index $
\ell' = 3\ell $.

\begin{figure}[!b]
    \centering
    \includegraphics[width=0.9\linewidth]{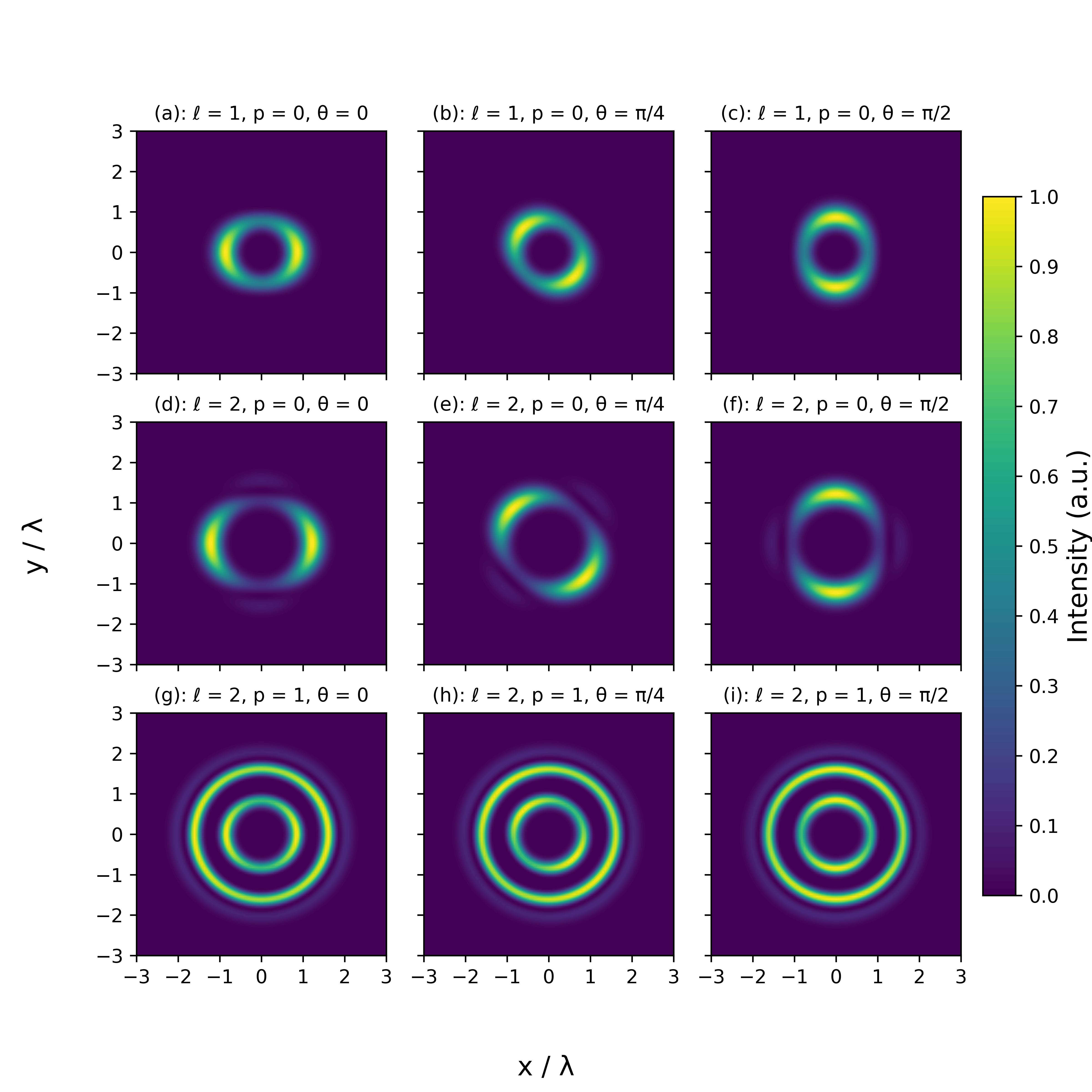}
    \caption{\small{Third-harmonic intensity distributions computed from~\eqref{eq:intensityAnswer} for varying $\ell$, $p$, and $\theta$. Each panel is normalized to its own maximum, with $\lambda = w_0$ throughout. Increasing $\ell$ broadens the ring structure, while increasing $p$ introduces additional radial nodes. The intensity is asymmetric, with enhanced lobes aligned along the input polarization direction, and the overall pattern rotates with $\theta$.}}
    \label{fig:THG3x3}
\end{figure}

We parameterize the linear polarization by an angle $\theta$ in the transverse plane such that
\begin{align}
\alpha &= \cos\theta, \\
\beta  &= \sin\theta,
\end{align}
where $\theta$ is measured with respect to the $x$-axis. This provides a continuous description of linear polarization states from $x$-polarization ($\theta=0$) through diagonal to $y$-polarization ($\theta=\pi/2$). In the expressions above the incident field retains its longitudinal component, which arises from the nonparaxial structure of the focused beam and plays a key role in enabling the nonlinear interaction. For the emitted harmonic field, however, we restrict attention to the transverse radiation components. Although longitudinal field components may be present locally in the focal region, the propagating third-harmonic radiation detected in the far field is overwhelmingly transverse, with longitudinal contributions strongly suppressed as the beam approaches the paraxial regime. Accordingly, the longitudinal component of the emitted field is neglected in what follows, and the harmonic polarization is taken to lie in the transverse plane. 
We obtain:
\begin{align}
|\mathbf{e}\cdot\mathbf{e}|^2
&=
\left[
1+\frac{1}{2k^2}\left(\frac{\ell^2}{r^2}-\gamma^2\right)
-\frac{1}{2k^2}\left(\gamma^2+\frac{\ell^2}{r^2}\right)\cos 2(\phi-\theta)
\right]^2
\nonumber \\ &+
\frac{\gamma^2 \ell^2}{k^4 r^2}\sin^2 2(\phi-\theta).
\label{eq:eedotegenlin}
\end{align}
and trivially \( |\mathbf{e}\cdot\overline{\mathbf{e}}'|^2 = 1 \). Clearly,~\eqref{eq:eedotegenlin} contains no terms odd in \(\ell\), so the intensity distribution in~\eqref{eq:intensityAnswer} is invariant under \(\ell \rightarrow -\ell\). The process therefore exhibits no chiral sensitivity, as expected on symmetry grounds. Substituting~\eqref{eq:eedotegenlin} into~\eqref{eq:intensityAnswer}, the resulting intensity distributions for various values of \(\theta\) are shown in Fig.~\ref{fig:THG3x3}. For \(\ell \neq 0\), the emitted harmonic exhibits vortex structure, while increasing \(p\) introduces radial nodes. The orientation of the input linear polarization gives rise to two clear features. Firstly, the harmonic field is not circularly symmetric, exhibiting distinct lobes of high intensity aligned parallel to the input polarization direction. This behaviour mirrors that observed in the linear intensity profile of tightly focused vortex beams~\cite{forbes2026vortex} and originates from the nonparaxial longitudinal field components. Secondly, the spatial intensity distribution rotates with the polarization angle \(\theta\), again consistent with the behaviour of the corresponding linear field.

Reversing the sign of \(\ell\) produces identical intensity distributions for fixed \(p\) and \(z\), confirming that the response depends only on \(|\ell|\) and is independent of the handedness of the vortex, in agreement with~\eqref{eq:eedotegenlin}.
For larger values of \(\ell\) and \(p\) [Fig.~\ref{fig:THG3x3}(d-i)], the intensity patterns become increasingly structured. Bright rings sharpen and dominate, reflecting the scaling of the signal as \((\tilde{f}_{LG})^6(\tilde{f}_{LG}')^2\), which enhances peak regions. 

These results indicate that the spatial structure of the THG field can be systematically controlled through the mode indices \(\ell\) and \(p\), providing a route to tailoring harmonic beam profiles for applications in imaging and spectroscopy.


\subsection{Evolution along $z$}

\begin{figure}[!t]
    \centering
    \includegraphics[width=0.9\linewidth]{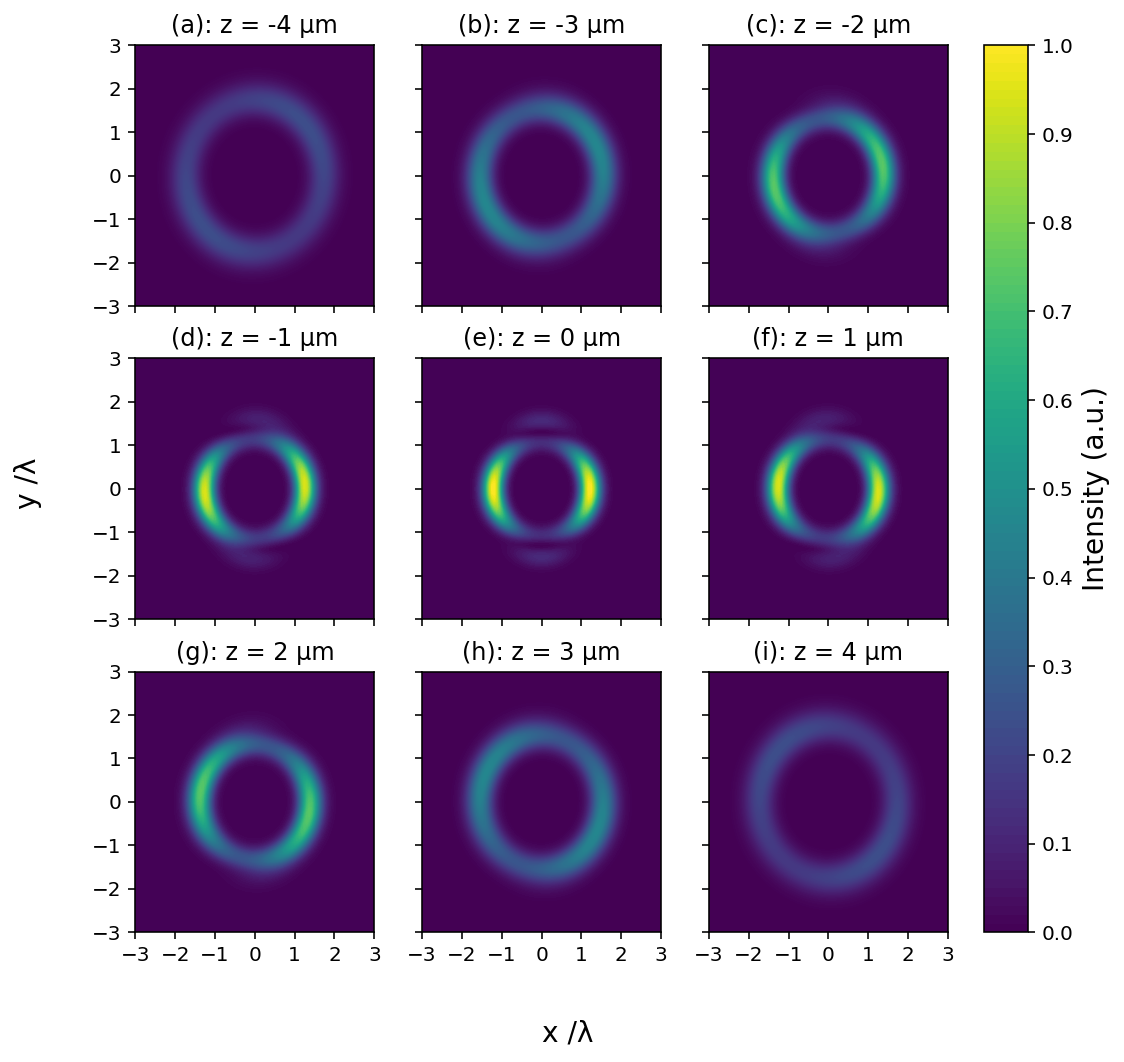}
    \caption{\small{Spatial intensity distributions along the direction of propagation for THG when $\ell=2$ and $p=0$, for an input beam with $\lambda=w_0$, and $x$-polarized incident and emission photons. The plots are normalized to the maximum of (e).}}
    \label{fig:THGzprop}
\end{figure}

Since these intensity distributions evolve along the propagation axis, the spatial dependence along the $z$-axis may also be considered. Taking subfigure \ref{fig:THG3x3}d as an example, Figure \ref{fig:THGzprop} shows variations along the direction of propagation. It is emphasized that these results are obtained from the local nonlinear source intensity, and therefore implicitly assume effective phase matching of the generated harmonic field.

Subfigure \ref{fig:THGzprop}e corresponds to subfigure \ref{fig:THG3x3}d. The remaining subfigures in Figure \ref{fig:THGzprop} show the evolution of the local intensity distribution along the $z$-axis. As $z$ increases, the structure becomes less intricate and tends towards an approximately symmetric ring. This behaviour is primarily governed by beam divergence and the changing balance between transverse and longitudinal field components. The apparent increase in radius reflects the natural spreading of the Laguerre-Gaussian mode.

Subfigures \ref{fig:THGzprop}e-\ref{fig:THGzprop}i exhibit an apparent rotation of the intensity pattern, observable by tracking the points of maximum intensity. Pairing subfigures as \ref{fig:THGzprop}a-\ref{fig:THGzprop}i, \ref{fig:THGzprop}b-\ref{fig:THGzprop}h, \ref{fig:THGzprop}c-\ref{fig:THGzprop}g, and \ref{fig:THGzprop}d-\ref{fig:THGzprop}f shows that reversing the sign of $z$ produces the same behaviour with opposite handedness, consistent with the symmetry of the focused field.

It is also noted that more intricate structure is most evident for $z \lesssim z_R$. The Rayleigh range, $z_R=\frac{\pi w_0^2}{\lambda}$, sets the characteristic axial scale of the beam, and in the present case reduces to $z_R=\pi w_0 \approx 3.14~\mu$m. Within this region, the field structure is dominated by the focal geometry.

Away from the focal plane, the Gouy phase introduces a mode-order-dependent axial phase mismatch between the nonlinear source and the generated harmonic field. As a result, coherent accumulation is reduced and the calculated intensity should be interpreted as a local source distribution rather than the fully propagated harmonic signal.

Accordingly, the discussion of Figure \ref{fig:THGzprop} is most directly applicable to experiments probing the focal region or within the Rayleigh range. Beyond this regime, the observed behaviour primarily reflects the evolution of the underlying beam structure, with reduced conversion efficiency expected due to incomplete phase matching.

\subsection{Circular polarization}

We now consider the case of circularly polarized excitation. The polarization coefficients may be written as
\begin{align}
\alpha=\frac{1}{\sqrt{2}}, \qquad \beta=\frac{i\sigma}{\sqrt{2}},
\end{align}
where $\sigma=\pm1$ denotes the helicity of the incident light. In contrast to the linear polarization case, the complex phase relation between the transverse field components corresponds to a nonzero spin angular momentum of $\sigma\hbar$ per photon.

For a purely transverse plane wave the scalar product $(\mathbf{e}\cdot\mathbf{e})$ vanishes for circular polarization, preventing third-harmonic generation in an isotropic medium \cite{craig1998molecular, tang1971selection, ford2018molecular, andrews1980harmonic}. However, for a focused beam the electric field possesses a longitudinal component arising from the nonparaxial structure of the mode. Substitution of the circular polarization coefficients into~\eqref{eq:8} shows that the surviving contribution to $(\mathbf{e}\cdot\mathbf{e})$ originates from this longitudinal field and carries an azimuthal phase dependence
\begin{align}
(\mathbf{e}\cdot\mathbf{e})
=
-\frac{1}{2k^{2}}\left(\gamma-\frac{\sigma\ell}{r}\right)^{2}\text{e}^{i2\sigma\phi}.
\end{align}

The nonlinear source term responsible for third-harmonic generation therefore carries the phase factor $\mathrm{e}^{i(3kz+(3\ell+2\sigma)\phi)}$. Phase matching then requires the emitted harmonic mode to possess corresponding longitudinal and azimuthal phase dependences with $k'=3k$ and $\ell'=3\ell+2\sigma$, ensuring conservation of energy and total angular momentum of the radiation field. The vortex charge of the generated harmonic is therefore
\begin{align}
\ell' = 3\ell + 2\sigma .
\end{align}

In the special case of a circularly polarized Gaussian input beam ($\ell=0$), the generated third harmonic consequently carries orbital angular momentum with $\ell' = 2\sigma$, demonstrating that tightly focused circularly polarized light can produce vortex harmonics even in isotropic media. For example, a tightly focused circularly polarized Gaussian beam can generate a third-harmonic field containing a vortex component with topological charge $\ell' = 2\sigma$, arising from spin-orbit interaction associated with the longitudinal field of the focused beam. This previously forbidden process therefore represents a further manifestation of spin-orbit interaction in light \cite{bliokh2015spin}.

\begin{figure}[!t]
    \centering
    \includegraphics[width=0.9\linewidth]{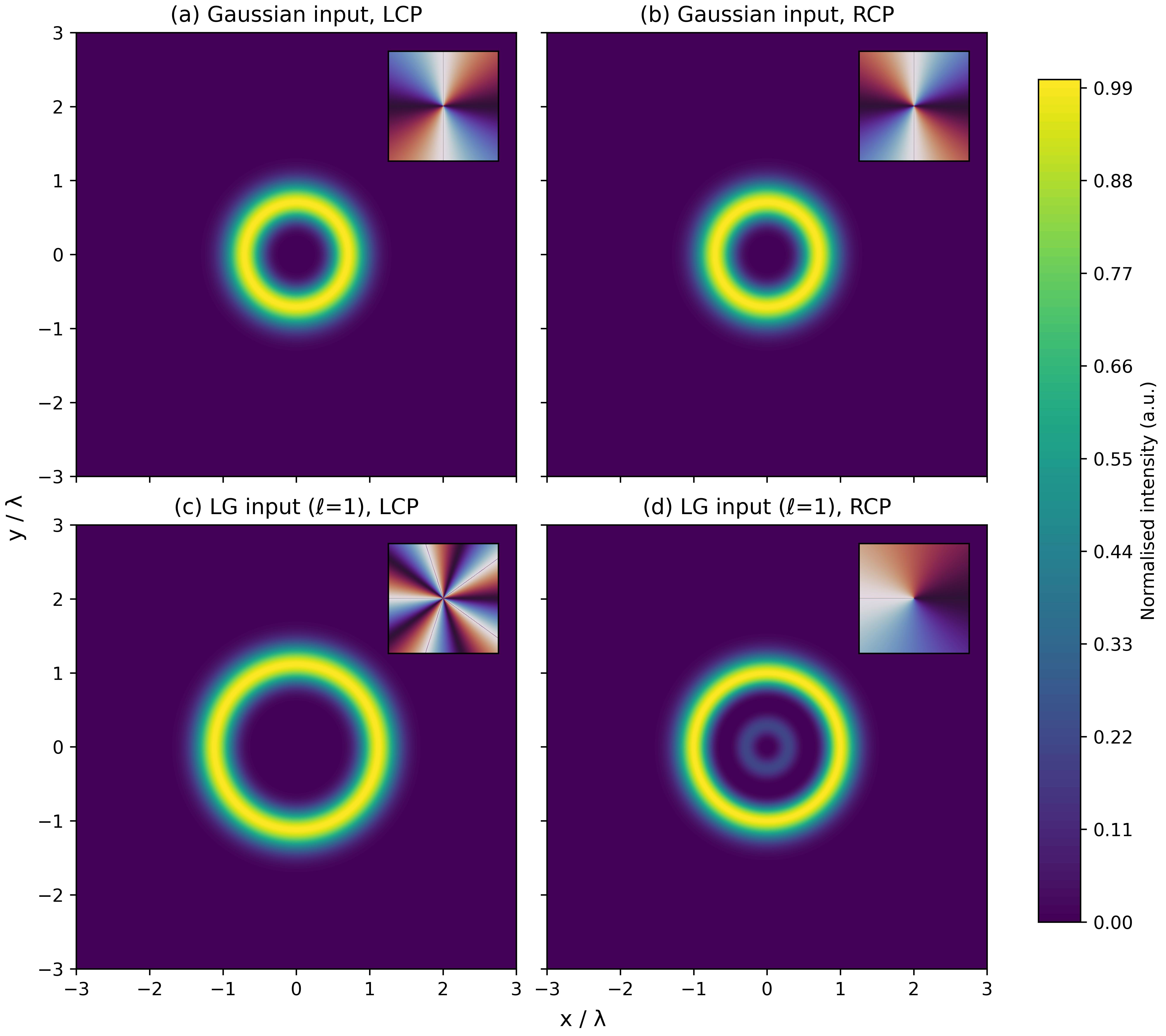}
    \caption{\small{Intensity of third-harmonic generation for circularly polarized input (phase distributions shown in insets). (a,b) For a Gaussian input ($\ell=0$), the generated harmonic carries a vortex of charge $\ell=\pm2$, with the sign set by the input helicity $\sigma$, consistent with angular momentum conservation. (c,d) For circularly polarized vortex beams, spin-orbit interaction emerges: parallel spin and OAM ($\sigma\ell>0$) produce a single-ring structure, whereas antiparallel spin and OAM ($\sigma\ell<0$) yield a distinct intensity profile. All panels are individually normalized, with $\lambda = w_0$.}}
    \label{fig:cpl}
\end{figure}

Again restricting attention to the transverse radiation components of the emitted harmonic field, we take $|\mathbf{e}\cdot\overline{\mathbf{e}}'|^2 = 1$, such that the remaining polarization dependence is fully contained in
\begin{align}
|\mathbf{e}\cdot\mathbf{e}|^2
=
\frac{1}{4k^{4}}
\left(\gamma-\frac{\sigma\ell}{r}\right)^{4}.
\end{align}
This quantity may then be substituted into~\eqref{eq:intensityAnswer} to model third-harmonic generation driven by tightly focused circularly polarized Laguerre-Gaussian modes. The results are shown in Fig.~\ref{fig:cpl}. 

Firstly, for a Gaussian input beam, we find that circular polarization generates a vortex harmonic: an input LCP beam produces \(\ell=+2\), while an input RCP beam produces \(\ell=-2\), demonstrating that the sign of \(\sigma\) determines the sign of the generated topological charge. For vortex input beams \((\ell \neq 0)\), we observe clear signatures of spin-orbit interaction \cite{bliokh2015spin}. In particular, the spatial structure of the harmonic depends on whether the spin and orbital angular momentum are parallel \(\big(\mathrm{sgn}\,\ell = \mathrm{sgn}\,\sigma\big)\) or anti-parallel \(\big(\mathrm{sgn}\,\ell \neq \mathrm{sgn}\,\sigma\big)\), leading to distinct intensity distributions. This behaviour is consistent with that observed in the linear regime for tightly focused vortex beams~\cite{forbes2026vortex}.

We emphasise that, for plane-wave light and paraxial beams with purely transverse fields, third-harmonic generation with circularly polarized input is forbidden \cite{andrews1980harmonic, tang1971selection, andrews2002optical}. The non-zero response obtained here arises entirely from the longitudinal field component introduced by focusing. This contribution scales with the degree of nonparaxiality and vanishes in the paraxial limit. 

\section{Third-Harmonic Vortex Dichroism}

In the previous treatment of THG, all photon interactions with the molecules were assumed to be purely electric-dipole in nature. Although the electric-dipole contributions in~(\ref{eq:Hint1}) are typically orders of magnitude larger than the corresponding magnetic-dipole terms, the latter enable chiral molecules to exhibit optical activity \cite{craig1998molecular, barron2009molecular}. In this section we show that a chiroptical third-harmonic generation vortex dichroism (THVD) arises from the interference between electric- and magnetic-dipole contributions to the nonlinear scattering amplitude, thereby introducing a sensitivity to the sign of the vortex topological charge in chiral media.

One important manifestation of linear optical activity is circular dichroism: the differential absorption of right- and left-handed circularly polarized light by chiral molecules. This phenomenon forms the basis of chiroptical spectroscopy, an incisive all-optical method for determining the absolute handedness of pharmaceuticals and other biologically relevant molecules \cite{polavarapu2016chiroptical, berova2011comprehensive, berova2012comprehensive}. As discussed in the previous section, under typical illumination conditions third-harmonic generation is forbidden for circularly polarized light in isotropic media \cite{tang1971selection, ford2018molecular, andrews1980harmonic}, and consequently no analogous THG chiroptical effect has previously been identified. In this section we show that optical activity can in fact arise in third-harmonic generation when vortex light is employed, leading to a new effect which we term third-harmonic vortex dichroism THVD. We emphasise that the purely electric-dipole result obtained in the previous section for circularly polarized input does not itself constitute a chiroptical effect.

The first step in deriving the scattered intensity for THVD involves modifying the time-ordered Feynman diagrams used for THG. Specifically, by replacing one electric dipole coupling with a magnetic transition, and repeating this process for all permutations of the photon interactions (see Supplementary Information), the matrix element is obtained as


\begin{align}
    M_{FI}
    &=-\left(\frac{\hbar c}{2 \epsilon_0V}\right)^2(k^3k')^{\frac{1}{2}}\sqrt{n(n-1)(n-2)} \nonumber \\ 
   &\times \sum\limits_\xi
    \overline{\tilde{f}'_{\mathrm{LG}}(r_\xi,z_\xi)}\,\tilde{f}_{\mathrm{LG}}^3(r_\xi,z_\xi)\,
    \mathrm{e}^{i\Delta\Phi(r_\xi,\phi_\xi,z_\xi)}
    \bigg[
    c\overline{e}_i'e_je_ke_l\beta_{ijkl}(\xi)
    \nonumber \\ &+ \overline{b}_i'e_je_ke_lG_{ijkl}(\xi)
    + \overline{e}_i'b_je_ke_lH_{ijkl}(\xi)
    + \overline{e}_i'e_jb_ke_lI_{ijkl}(\xi)
    \nonumber \\ & + \overline{e}_i'e_je_kb_lJ_{ijkl}(\xi)
    \bigg].
    \label{eq:dichroismMatrix}
\end{align}


where $\Delta\Phi(r_\xi,\phi_\xi,z_\xi)$ is the phase mismatch defined in~(\ref{phase}), evaluated at molecular centre $\xi$. $b_i$ represents the magnetic field vector of photon $i$. $G_{ijkl}$, $H_{ijkl}$, $I_{ijkl}$, and $J_{ijkl}$ are purely imaginary fourth-rank tensors corresponding to the light-matter interaction pathways illustrated in the Supplementary Information. The full expressions for these tensors are provided in Supplementary Information. Using  (\ref{eq:dichroismMatrix}), and exploiting similarities between the $H_{ijkl}$, $I_{ijkl}$, and $J_{ijkl}$ tensors, the rotationally averaged intensity distribution is given by 
\begin{align}
    I_\text{E1+M1}=&\frac{9\overline{I}_0^3g^{(3)}k^4N_{\mathrm{eff}}^{\,2}}{6400\pi^2\varepsilon_0^4c^4} \tilde{f}^6_{LG}\tilde{f}'^2_{LG} \bigg| 6c(\textbf{e}\cdot\textbf{e})(\textbf{e}\cdot\overline{\textbf{e}}')\beta_{\lambda\lambda\mu\mu} \nonumber \\ &+ 6(\textbf{e}\cdot\textbf{e})(\textbf{e}\cdot\overline{\textbf{b}}')G_{\lambda\lambda\mu\mu} \nonumber \\ & + (\textbf{e}\cdot\textbf{e})(\textbf{b}\cdot\overline{\textbf{e}}') \bigg(H_{\lambda\lambda\mu\mu} +3H_{\lambda\mu\lambda\mu} +2H_{\lambda\mu\mu\lambda}\bigg) \nonumber \\ &+ (\textbf{b}\cdot\textbf{e})(\textbf{e}\cdot\overline{\textbf{e}}') \bigg(7H_{\lambda\lambda\mu\mu}+H_{\lambda\mu\lambda\mu} +4H_{\lambda\mu\mu\lambda}\bigg)\bigg|^2
    \label{eq:dichroismIntensity}
\end{align}

Further details of the rotational average can be found in the Supplementary Information. The first term in the vertical bars in~(\ref{eq:dichroismIntensity}) is simply the electric-dipole contribution to the THG scattering intensity from the previous section. It is established that the origin of natural optical activity arises from the interferences between electric dipole couplings (E1) with both magnetic dipole (M1) and electric quadrupole (E2) couplings \cite{andrews2018quantum}. The orientational average of the fifth-rank tensor associated with the E1E2 interaction vanishes for an isotropic molecular ensemble. This follows because no isotropic tensor of odd rank can be formed from Kronecker delta invariants, and hence the rotational average of the E1E2 contribution is identically zero. This justifies why we have concentrated solely on dipole coupling.  

Therefore only E1M1 contributions will be considered, justifying the earlier truncation of the multipolar interaction Hamiltonian to dipolar coupling. To identify the terms responsible for third-harmonic optical activity, the relevant cross terms with an overall odd parity must be isolated. Electric-dipole interactions have odd parity, whereas magnetic-dipole interactions have even parity. Consequently, the tensor $\beta_{\lambda\lambda\mu\mu}$, containing four electric-dipole interactions, possesses overall even parity, while the remaining tensors acquire odd parity through the introduction of a single magnetic-dipole interaction. The chiroptical response therefore arises from terms containing one $\mathbf{b}$ vector, since an overall odd parity to leading order requires an odd number of electric-field factors combined with a single magnetic-field contribution. In the leading order chiral contribution six distinct contributing terms remain, which can be grouped into three pairs:
\begin{align}
    I_\text{THVD}=&\frac{27\overline{I}_0^3g^{(3)}k^4N_{\mathrm{eff}}^{\,2}}{1600\pi^2\varepsilon_0^4c^3}\beta_{\lambda\lambda\mu\mu} f^6_{LG}f'^2_{LG} \textup{Re}\bigg[] |\textbf{e}\cdot\textbf{e}|^2(\overline{\textbf{e}}\cdot\textbf{e}')(\textbf{e}\cdot\overline{\textbf{b}}')G_{\lambda\lambda\mu\mu} \nonumber \\ & + |\textbf{e}\cdot\textbf{e}|^2(\overline{\textbf{e}}\cdot\textbf{e}')(\textbf{b}\cdot\overline{\textbf{e}}') \Big(H_{\lambda\lambda\mu\mu} + 3H_{\lambda\mu\lambda\mu}+2H_{\lambda\mu\mu\lambda}\Big) \nonumber \\ & + |\textbf{e}\cdot\overline{\textbf{e}}'|^2(\overline{\textbf{e}}\cdot\overline{\textbf{e}})(\textbf{b}\cdot\textbf{e}) \Big(7H_{\lambda\lambda\mu\mu} +H_{\lambda\mu\lambda\mu}+4H_{\lambda\mu\mu\lambda}\Big) \bigg]
    \label{eq:dichroismIntensityApproximation3} 
\end{align}

In the previous treatment of THG all contributing terms depended solely on the electric-field polarization vectors $\mathbf{e}$. Since electric-dipole interactions possess odd parity, combinations involving only electric fields yield an overall even parity and therefore cannot give rise to chiral effects. In~(\ref{eq:dichroismIntensityApproximation3}), however, terms appear that contain magnetic-field vectors $\mathbf{b}$, enabling optical activity through electric-magnetic dipole interference.

\subsection{Linear polarization}

We now focus on linearly polarized input beams. In this case the circular-polarization helicity $\sigma$ vanishes, removing any chirality associated with the spin angular momentum of the field. Consequently, any non-zero chiroptical signal must originate solely from the handedness of the optical wavefront, characterized by the pseudoscalar vortex charge $\ell$.

We once again parameterize the incident linearly polarized field by an angle \(\theta\) in the transverse plane as $\mathbf{e} = \cos\theta\,\hat{\mathbf{x}} + \sin\theta\,\hat{\mathbf{y}},$ with the corresponding magnetic field orthogonal to \(\mathbf{e}\) and also confined to the transverse plane. As in the preceding THG analysis, we restrict attention to the transverse radiation components of the emitted harmonic field, neglecting the longitudinal component that arises locally from the nonlinear source but is strongly suppressed in the far-field propagation. The detected harmonic is taken to share the same transverse polarization as the incident field, so that \(\overline{\mathbf{e}}'\parallel \mathbf{e}\) and \(\overline{\mathbf{b}}'\) remains orthogonal to \(\overline{\mathbf{e}}'\). Under these conditions the mixed contractions \(\mathbf{e}\cdot\overline{\mathbf{b}}'\) and \(\mathbf{b}\cdot\overline{\mathbf{e}}'\) vanish by orthogonality, leaving the only surviving electric-magnetic contribution proportional to $(\mathbf{b}\cdot\mathbf{e})(\overline{\mathbf{e}}\cdot\overline{\mathbf{e}})$.

The magnetic-field vectors appearing in~(\ref{eq:dichroismIntensityApproximation3}) are given by
\begin{align}
    \textbf{b}(\textbf{r})&=\alpha \mathbf{\hat{y}}-\beta \mathbf{\hat{x}}+ \nonumber \\ & \mathbf{\hat{z}}\frac{i}{k}\bigg\{\alpha\bigg(\gamma \sin\phi + \frac{i \ell }{r} \cos\phi\bigg)-\beta \bigg(\gamma \cos\phi - \frac{i \ell }{r} \sin\phi\bigg) \bigg\}
\end{align}
and
\begin{align}
    \overline{\textbf{b}}'(\textbf{r})=\overline{\alpha}' \mathbf{\hat{y}}-\overline{\beta}' \mathbf{\hat{x}}
\end{align}

Using these vectors yields:
\begin{align}
(\mathbf{b}\cdot\mathbf{e})
&=
\frac{1}{k^2}
\Bigg[
\left(\gamma^2+\frac{\ell^2}{r^2}\right)
\left(
\alpha\beta\cos 2\phi
-\frac{1}{2}(\alpha^2-\beta^2)\sin 2\phi
\right)
\nonumber \\ & +
i\frac{\gamma\ell}{r}
\left(
(\beta^2-\alpha^2)\cos 2\phi
-2\alpha\beta\sin 2\phi
\right)
\Bigg].
\label{eq:b.e}
\end{align}
It is noteworthy that the transverse electric and magnetic field components are identically orthogonal for arbitrary Jones coefficients $\alpha$ and $\beta$, as required by Maxwell’s equations for a propagating beam. Consequently $(\mathbf{b}\cdot\mathbf{e})$ receives contributions only from the longitudinal field components associated with the nonparaxial structure of the beam.

The next step is to combine all polarization inner products appearing in the non-zero term of~\eqref{eq:dichroismIntensityApproximation3}. Assuming again that the emitted harmonic is probed in the far field, we have trivially \( |\mathbf{e}\cdot\overline{\mathbf{e}}'|^2=1 \), leaving the product \( (\overline{\mathbf{e}}\cdot\overline{\mathbf{e}})(\mathbf{b}\cdot\mathbf{e}) \). For a general linear polarization state, parameterized by the angle \(\theta\), this quantity can be written compactly in terms of the relative azimuthal coordinate \(\phi-\theta\). At the focal plane \(z=0\), where \(\gamma\) is real, we obtain
\begin{align}
&(\mathbf{b}\cdot\mathbf{e})(\overline{\mathbf{e}}\cdot\overline{\mathbf{e}})
=
-\frac{1}{k^2}
\left[
\frac{1}{2}\left(\gamma^2+\frac{\ell^2}{r^2}\right)\sin 2(\phi-\theta)
+i\gamma\frac{\ell}{r}\cos 2(\phi-\theta)
\right] \nonumber\\
&\quad\times
\left[
1+\frac{1}{k^2}
\left(
\frac{\ell^2}{r^2}\sin^2(\phi-\theta)
-\gamma^2\cos^2(\phi-\theta)
-i\gamma\frac{\ell}{r}\sin 2(\phi-\theta)
\right)
\right].
\end{align}

\noindent
Here \(\theta\) denotes the angle of the input linear polarization with respect to the \(x\)-axis, such that \(\theta=0\) corresponds to \(x\)-polarization, \(\theta=\pi/4\) to diagonal polarization, and \(\theta=\pi/2\) to \(y\)-polarization.

The observable intensity is determined by the real part of the field-molecular product. The tensor combination
\[
7H_{\lambda\lambda\mu\mu}+H_{\lambda\mu\lambda\mu}+4H_{\lambda\mu\mu\lambda}
= i\,\mathcal{H},
\qquad \mathcal{H}\in\mathbb{R},
\]
is purely imaginary, as expected for magnetic-dipole contributions. The relevant term therefore reduces to the imaginary part of the field factor,
\begin{align}
\Im\!\big[(\overline{\mathbf e}\cdot\overline{\mathbf e})(\mathbf b\cdot\mathbf e)\big]
&=
\frac{\gamma\ell}{k^2 r}\cos 2(\phi-\theta)
\nonumber \\ &+
\frac{\gamma\ell}{k^4 r}
\left[
\frac{\ell^2}{r^2}\cos^2(\phi-\theta)
+
\gamma^2\sin^2(\phi-\theta)
\right],
\label{eq:Im(e.e)(b.e)}
\end{align}
The linear dependence on \(\ell\) shows that the signal reverses with the handedness of the optical vortex. The resulting THVD intensity is
\begin{align}
&I_{\mathrm{THVD}}^{\mathrm{lin}}
=\;
\frac{27\overline{I}_0^3g^{(3)}k^4N_{\mathrm{eff}}^{\,2}}{1600\pi^2\varepsilon_0^4c^3}\,
\beta_{\lambda\lambda\mu\mu}\,
f_{\mathrm{LG}}^6 f_{\mathrm{LG}}'^2\,
\mathcal{H}
\nonumber\\
&\times
\left[
\frac{\gamma\ell}{k^2 r}\cos 2(\phi-\theta)
+
\frac{\gamma\ell}{k^4 r}
\left(
\frac{\ell^2}{r^2}\cos^2(\phi-\theta)
+
\gamma^2\sin^2(\phi-\theta)
\right)
\right].
\label{eq:THVD_lin_final_general}
\end{align}

\begin{figure}[!b]
    \centering
    \includegraphics[width=0.78\linewidth]{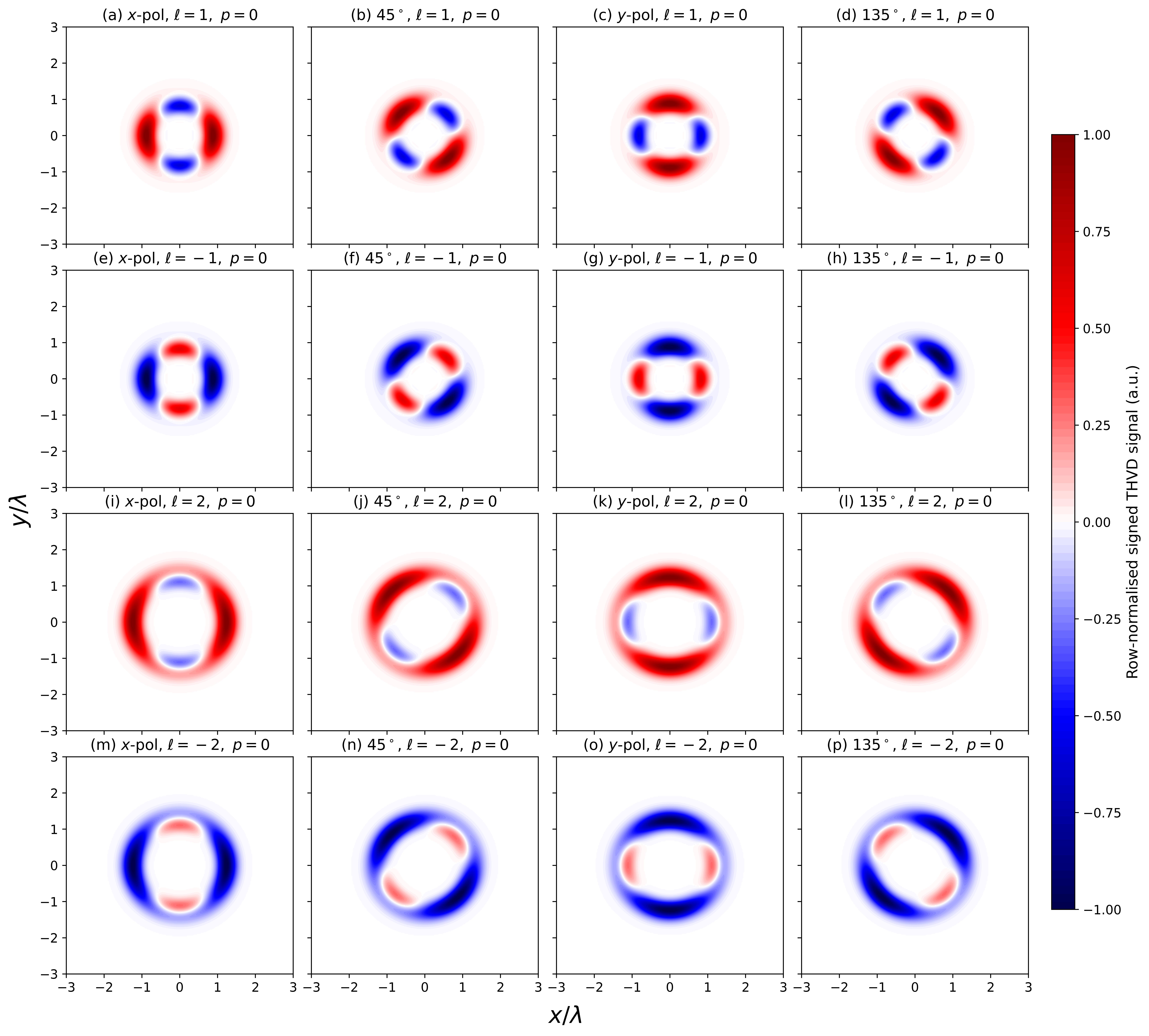}
    \caption{\small{Spatial distribution of the THVD signal computed from~\eqref{eq:THVD_lin_final_general} for varying $\ell$ and $\theta$. The sign of the distribution reverses with the handedness of $\ell$ and rotates with $\theta$. In all cases the spatially integrated signal is non-zero, with the effect becoming more pronounced for larger $\ell$. All panels are individually normalized, with $\lambda = w_0$.}}
    \label{fig:THVD6x3}
\end{figure}

Using~(\ref{eq:THVD_lin_final_general}) as the framework to model THVD in fluids, Figure \ref{fig:THVD6x3} illustrates the spatial distribution in the focal plane of this process. For clarity, all molecular response tensors are taken to be equal in magnitude in the simulations presented here for illustrative purposes, and this assumption is maintained throughout the manuscript. A fully quantitative treatment for a specific system would require evaluation of these tensors using quantum chemical methods, which is beyond the scope of this work. Since all contributing tensor terms enter with the same polarization structure under the conditions considered here (far-field detection of the transverse harmonic field), their relative magnitudes do not alter the qualitative form of the spatial distributions, but only rescale the overall signal. The results therefore capture the general, field-driven features of the response, governed by symmetry and the structure of the light-matter interaction rather than by molecule-specific tensor values.


The spatial distributions shown in Fig.~\ref{fig:THVD6x3} represent the signed THVD signal for varying input polarization angle \(\theta\) and topological charge \(\ell\), with \(p=0\). A prominent feature is the appearance of alternating positive and negative lobes arranged azimuthally about the beam axis. A defining property of these distributions is their odd dependence on \(\ell\). Comparison of rows corresponding to \(\ell\) and \(-\ell\) shows identical spatial structure with complete sign reversal, such that regions of positive and negative signal are interchanged. This confirms that the THVD signal is linear in \(\ell\), and therefore directly encodes the handedness of the optical vortex. In contrast to the achiral THG intensity, which depends only on \(|\ell|\), the THVD response is intrinsically sensitive to the sign of the topological charge.

The orientation of the lobes is set by the input polarization angle $\theta$. Varying $\theta$ rotates the entire THVD pattern, consistent with the angular dependence in~\eqref{eq:THVD_lin_final_general}. A similar angle-dependent response has been reported in two-photon absorption of vortex beams~\cite{cheeseman2025nonlinear}, in contrast to the circular symmetry typical of the linear regime. Increasing \(|\ell|\) modifies the spatial structure while preserving its symmetry. For \(|\ell|=1\), the signal consists of broader lobes, whereas for \(|\ell|=2\) the distribution becomes more sharply defined and ring-like, with enhanced contrast at larger radii. This reflects the underlying Laguerre-Gaussian mode structure, with the THVD signal inheriting both the radial redistribution of intensity and the azimuthal phase structure.

The resulting patterns are symmetric in magnitude but antisymmetric in sign, consistent with the pseudoscalar character of the E1M1 term. Nodal lines separating regions of opposite sign correspond to directions along which the chiral contribution vanishes. Overall, these results demonstrate that THVD provides a spatially resolved chiral observable in isotropic media, in which the vortex handedness determines the sign of the response while the input polarization controls its orientation.

Unlike linear chiroptical measures, nonlinear optical signals do not admit a conserved electromagnetic pseudoscalar analogous to optical chirality \cite{bliokh2011characterizing, cameron2017chirality}. Optical chirality arises from the duality symmetry of Maxwell’s equations and is quadratic in the fields, such that for linearly polarized beams its spatial integral vanishes even in the presence of longitudinal field components \cite{forbes2026vortex, forbes2026twisted}.

In contrast, the present third-harmonic vortex dichroism signal originates from higher-order field-matter interactions, involving products of three electric fields and one magnetic field. The resulting pseudoscalar is therefore not a property of the electromagnetic field alone, but of the combined light-matter interaction. As a consequence, there is no symmetry constraint enforcing cancellation of the spatial integral. The nonlinear interaction effectively converts the structured, but globally achiral, field into a non-vanishing pseudoscalar observable. Accordingly, the integrated signal can remain finite even for linearly polarized input, with its sign determined by the vortex charge $\ell$.

\subsection{Circular polarization}

Having established that a non-zero THVD signal can arise even in the absence of helicity $\sigma$ associated with polarization, we now consider circularly polarized excitation. In contrast to the linearly polarized case examined above, circular polarization introduces an intrinsic electromagnetic handedness associated with the spin angular momentum of the field. The resulting chiroptical response therefore reflects a nonparaxial coupling between the spin helicity $\sigma$ and the orbital angular momentum of the vortex beam characterized by the topological charge $\ell$. In this regime the THVD signal is no longer determined solely by the wavefront structure of the optical vortex, but instead arises from the coupling of the circular polarization state to the azimuthal phase structure via the longitudinal field components. This spin-orbit interaction is encoded through the combination $(\gamma-\sigma\ell/r)$, which modifies the spatial weighting of the signal.

\begin{align}
&I^{\mathrm{circ}}_\text{THVD}
=\;
\frac{27\overline{I}_0^3g^{(3)}N_{\mathrm{eff}}^{\,2}}{6400\pi^2\varepsilon_0^4c^3}\,
\beta_{\lambda\lambda\mu\mu}\,
\tilde{f}_{\mathrm{LG}}^6 \tilde{f}_{\mathrm{LG}}'^2\,
\delta_{\sigma\sigma'}\, \nonumber \\ & \times
\mathrm{Re}\!\Bigg[
i\sigma
\left(
\gamma-\frac{\sigma\ell}{r}
\right)^4
\Big(
G_{\lambda\lambda\mu\mu}
-8H_{\lambda\lambda\mu\mu}
-4H_{\lambda\mu\lambda\mu}
-6H_{\lambda\mu\mu\lambda}
\Big)
\Bigg].
\label{eq:THVD_circ_final}
\end{align}

\begin{figure}[!t]
    \centering
    \includegraphics[width=0.9\linewidth]{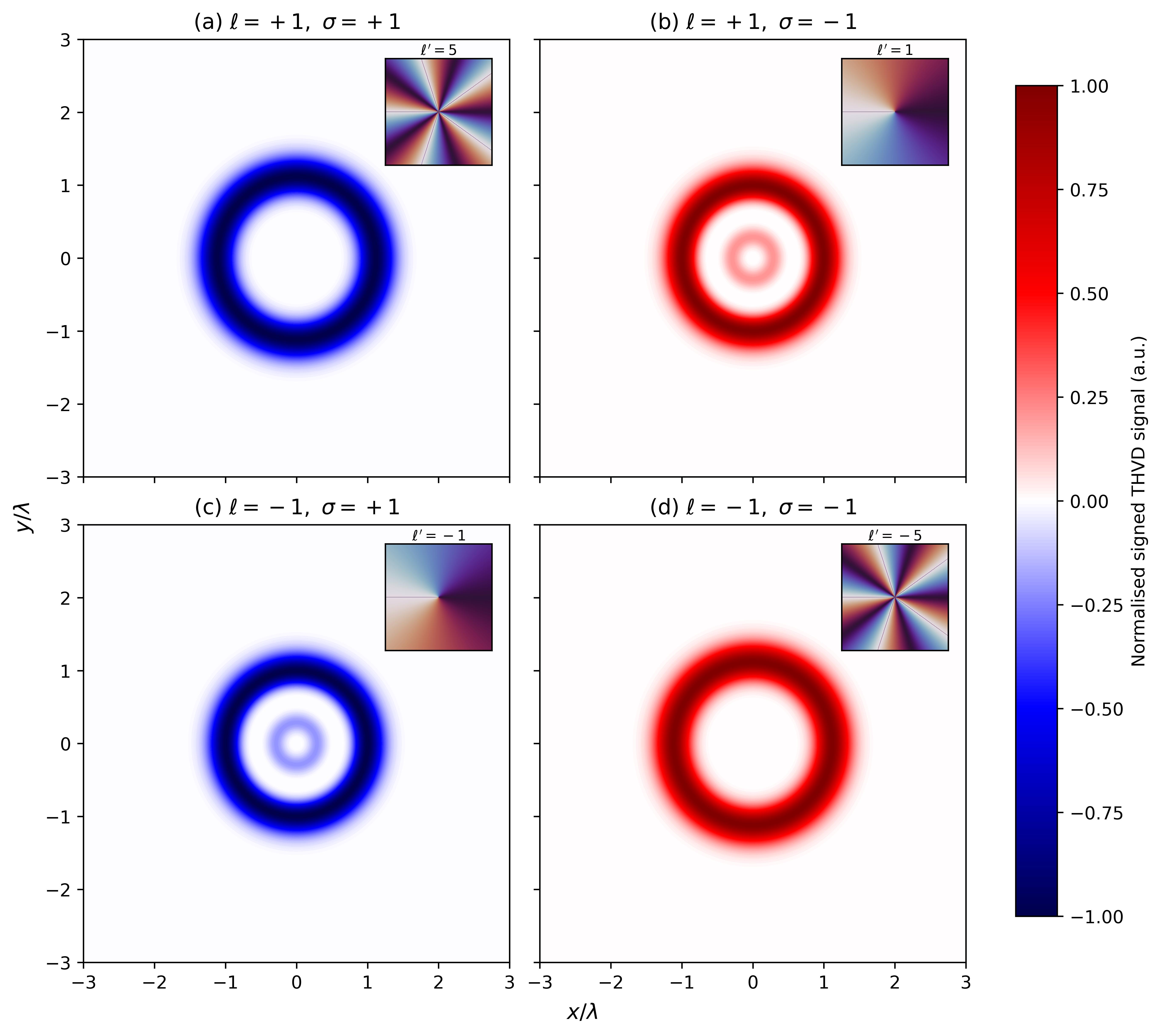}
    \caption{\small{Intensity of the circularly polarized THVD signal computed from~\eqref{eq:THVD_circ_final}. The response is governed by nonparaxial spin-orbit interaction between the helicity $\sigma$ and vortex charge $\ell$. The overall sign is determined primarily by the input circular polarization, while the value and sign of $\ell$ modify the radial profile of the signal. Unlike the linearly polarized case, the distributions are radially symmetric. All panels are individually normalized, with $\lambda = w_0$.}}
    \label{fig:cplTHVD}
\end{figure}

The spatial distributions corresponding to~\eqref{eq:THVD_circ_final} are shown in Fig.~\ref{fig:cplTHVD}. In contrast to the linearly polarized case, the THVD signal exhibits a clear spin-orbit interaction between $\sigma$ and $\ell$. The global sign of the response is determined primarily by the helicity $\sigma$, while the vortex charge $\ell$ modifies the spatial structure through the radial factor $(\gamma-\sigma\ell/r)^4$. As a result, changing the sign of $\ell$ does not simply invert the signal, but instead reshapes the radial profile through the spin-orbit interaction.

A further striking feature is the emergence of radially symmetric ring-like intensity profiles. Unlike the azimuthally structured patterns observed for linear polarization, the circularly polarized case produces distributions that are independent of the azimuthal angle \(\phi\). This reflects the absence of a preferred transverse direction in circular polarization, with the spin angular momentum imposing rotational symmetry on the response. The THVD signal is therefore primarily encoded in the radial dependence of the field, rather than in angular modulation.

The influence of spin-orbit interaction is evident when comparing different combinations of $\sigma$ and $\ell$. For a given helicity $\sigma$, varying the sign and magnitude of $\ell$ redistributes the intensity radially: some configurations produce a single dominant ring, whereas others generate additional inner structure. This behaviour is consistent with the form of~\eqref{eq:THVD_circ_final}, in which the factor $(\gamma-\sigma\ell/r)^4$ modifies the radial weighting of the signal through the nonparaxial interaction of spin helicity and orbital angular momentum. Reversing both the spin helicity \(\sigma\) and the vortex charge \(\ell\) produces a complete inversion of the THVD signal while preserving the spatial structure. This spin-orbit behaviour mirrors that of CPL-driven THG in Figure~\ref{fig:cpl}.

Overall, these results show that circularly polarized excitation leads to a qualitatively different THVD regime, in which the response is governed by the interplay of spin and orbital angular momentum and manifests predominantly through radially structured, sign-sensitive intensity distributions.

\section{Discussion and Conclusion}\label{Conclusion}
Third-harmonic generation in isotropic molecular fluids is a non-chiroptical process, dominated by the electric-dipole \(\chi^{(3)}\) response and insensitive to molecular handedness \cite{andrews2002optical, craig1998molecular}. Moreover, for circularly polarized plane waves the process is angular-momentum forbidden, so that no conventional third-harmonic analogue of circular dichroism has been available in such media \cite{andrews1980harmonic, tang1971selection}. The present work shows that both of these limitations can be overcome by structured light. In particular, the nonparaxial longitudinal field components of focused Laguerre-Gaussian beams permit third-harmonic generation with circularly polarized excitation in an isotropic fluid, establishing a mechanism that is unavailable for plane-wave or purely transverse fields. This significantly broadens the known symmetry landscape of THG in molecular media. A closely related effect was recently demonstrated in an optical nanofibre, where confinement-induced photonic spin-orbit interaction allows a circularly polarized Gaussian pump to drive third-harmonic generation into an optical vortex mode that would be forbidden in an isotropic bulk medium under conventional conditions \cite{ha2025optical}. In that case, the key physics is that subwavelength confinement generates spin–orbit-entangled eigenmodes, so that total angular momentum remains the relevant conserved quantity even though spin and orbital angular momenta are no longer separately conserved. The present results point to the same underlying origin, namely spin-orbit interaction, but in our case this arises through the structured and tightly focused optical field itself rather than through nanofibre-guided mode confinement. Indeed, third-harmonic generation from circularly polarised light has previously been observed under tight focusing, where the effect was attributed to depolarisation of the field \cite{dharmadhikari2015generation}. While this was not framed in terms of optical angular momentum, the present results suggest that such observations can be understood more generally as manifestations of spin–orbit interaction. Taken together, these studies show that the apparent lifting of conventional CPL-based selection rules in THG is a genuine signature of spin-orbit coupling, and not something unique to one specific platform.

Beyond this, we have shown that optical vortices enable the first chiroptical analogue of third-harmonic generation in isotropic molecular fluids. Through electric-dipole-magnetic-dipole interference, the generated third-harmonic signal acquires a contribution that is odd in the vortex charge and reverses sign upon inversion of either $\ell$ or the molecular handedness. This third-harmonic vortex dichroism therefore constitutes, to our knowledge, the first viable mechanism for chiral discrimination in THG from an isotropic fluid. In contrast to conventional chiroptical effects based on circular polarization, this regime can be driven by the handedness of the wavefront itself, with orbital angular momentum providing a key control parameter. We also identify a complementary regime in which a more conventional chiroptical THG signal arises from circularly polarized input, governed primarily by the helicity $\sigma$.

A notable consequence is that chiral-sensitive third-harmonic signals can be generated in randomly oriented media using both linearly polarized and circularly polarized vortex beams. The former represents a qualitatively distinct operating regime from standard circular-dichroism-based methods, emerging from the combined presence of longitudinal field components, transverse inhomogeneity, and geometric chirality in structured light. The sign and spatial form of the THVD signal can in principle be tuned through the vortex indices, suggesting a route to mode-resolved nonlinear chiroptical spectroscopy. Chiroptical nonlinear techniques are an active area of research, particularly in second- and third-order harmonic scattering in molecules and nanostructures \cite{shelton2018third, collins2019first, verreault2019hyper, ohnoutek2021optical, ohnoutek2022third, andrews2020irreducible}. Our results further indicate that tightly focused circularly polarized beams can generate a coherent chiroptical THG response, which may provide insight into the origin of previously reported incoherent nonlinear chiroptical signals.

The spatial structure of the predicted signals also points to possible applications beyond bulk spectroscopy. Since THG is intrinsically sensitive to local field structure and often strongest near boundaries and regions of optical inhomogeneity, vortex-driven THG and THVD may furnish new forms of spatially resolved nonlinear imaging in molecular fluids, soft matter, and interfacial environments. In such settings, the combination of harmonic contrast with sensitivity to vortex handedness could provide information not accessible to standard THG, while the circularly polarized THG channel identified here offers an additional means of controlling the generated harmonic mode through spin-orbit interaction.

Overall, these results show that structured light does more than modify the spatial profile of third-harmonic generation: it changes which nonlinear optical processes are allowed in isotropic molecular media. Focused optical vortices permit both the generation of third harmonics with circularly polarized excitation, previously deemed forbidden, and the emergence of a genuinely chiroptical third-harmonic response. Together, these findings expand the foundations of nonlinear molecular optics and identify orbital angular momentum and spin-orbit interaction as powerful degrees of freedom for controlling and probing harmonic generation in fluid media.

\begin{backmatter}

\bmsection{Disclosures} The authors declare no conflicts of interest.

\bmsection{Data availability} No data were generated or analyzed in the presented research.

\bmsection{Supplementary Information} See Supplementary Information for supporting content.
\end{backmatter}

\bibliography{sample}

\bibliographyfullrefs{sample}


\ifthenelse{\equal{\journalref}{aop}}{%
\section*{Author Biographies}
\begingroup
\setlength\intextsep{0pt}
\begin{minipage}[t][6.3cm][t]{1.0\textwidth} 
  \begin{wrapfigure}{L}{0.25\textwidth}
    \includegraphics[width=0.25\textwidth]{john_smith.eps}
  \end{wrapfigure}
  \noindent
  {\bfseries John Smith} received his BSc (Mathematics) in 2000 from The University of Maryland. His research interests include lasers and optics.
\end{minipage}
\begin{minipage}{1.0\textwidth}
  \begin{wrapfigure}{L}{0.25\textwidth}
    \includegraphics[width=0.25\textwidth]{alice_smith.eps}
  \end{wrapfigure}
  \noindent
  {\bfseries Alice Smith} also received her BSc (Mathematics) in 2000 from The University of Maryland. Her research interests also include lasers and optics.
\end{minipage}
\endgroup
}{}

\end{document}